# Stephenson et al.'s ecological fallacy

Eastaugh CS[*a], Thurnher C[b], Hasenauer H[b] and Vanclay JK[a]

a   Forest Research Centre, School of Science, Environment and Engineering, Southern Cross University, PO Box 157, Lismore, NSW 2480, Australia.

b  Institute of Silviculture, University of Natural Resources and Life Sciences (BOKU), Peter-Jordan Strasse 82, 1190 Vienna, Austria.

* chris.eastaugh@scu.edu.au

After more than a century of research the typical growth pattern of a tree was thought to be fairly well understood. Following germination height growth accelerates for some time, then increment peaks and the added height each year becomes less and less. The cross sectional area (basal area) of the tree follows a similar pattern, but the maximum basal area increment occurs at some time after the maximum height increment. An increase in basal area in a tall tree will add more volume to the stem than the same increase in a short tree, so the increment in stem volume (or mass) peaks very late. Stephenson et al. challenge this paradigm, and suggest that mass increment increases continuously. Their analysis methods however are a textbook example of the 'ecological fallacy', and their conclusions therefore unsupported.

The ecological fallacy[1] was most famously described by the sociologist William Robinson (1913-1996)[1], who pointed out that correlations present in aggregated data do not imply that the same correlations are present in the individuals[3]. In the context of the Stephenson et al. paper[4], this means that the fact that large trees of a species tend to have higher mass increment than small trees in no way implies that increment in individuals is a monotonic function of tree size. Robinson gave a mathematical proof of the fallacy, but we can demonstrate the effect here using tree measurements pertinent to Stephenson et al.'s study.

Adolf Ritter von Guttenberg (1839-1917) compiled one of the classic datasets of forestry science[5,6]. In 95 stands of Norway Spruce (*Picea abies* [L] Karst) across the Austrian Alps he felled 107 average trees and cut the stems into sections of between 1.0 and 4.0 metres length. The volume of each section was calculated individually and summed to give a total stem volume. By counting and measuring the annual growth rings in each section he was able to record a quite detailed growth history for each tree. We use here the data from the Hinterberg forest region in Styria, and select the six trees taken from Site Class 1 plots (the most productive). Tree ages at felling were 140 and 150 years. Aboveground biomass (including branches and foliage) is derived from the volume, diameter and height data using relationships from the standard Austrian biomass functions[7].

When we aggregate the data for these six trees and fit a piecewise linear regression (using Stephenson et al.'s software scripts[8]) we see that the results do indeed suggest monotonic mass accumulation (fig 1a). However, when we look at the course of mass increment for the individual trees (fig 1b), we see that typically growth accelerates while the tree is young or maturing, then levels and falls as maturity turns to senescence[9]. While Stephenson et al.'s analysis can tell us that the growth rate of large trees is typically greater than the growth rate of small trees, it does not follow that any individual tree will have a continuously increasing growth rate.

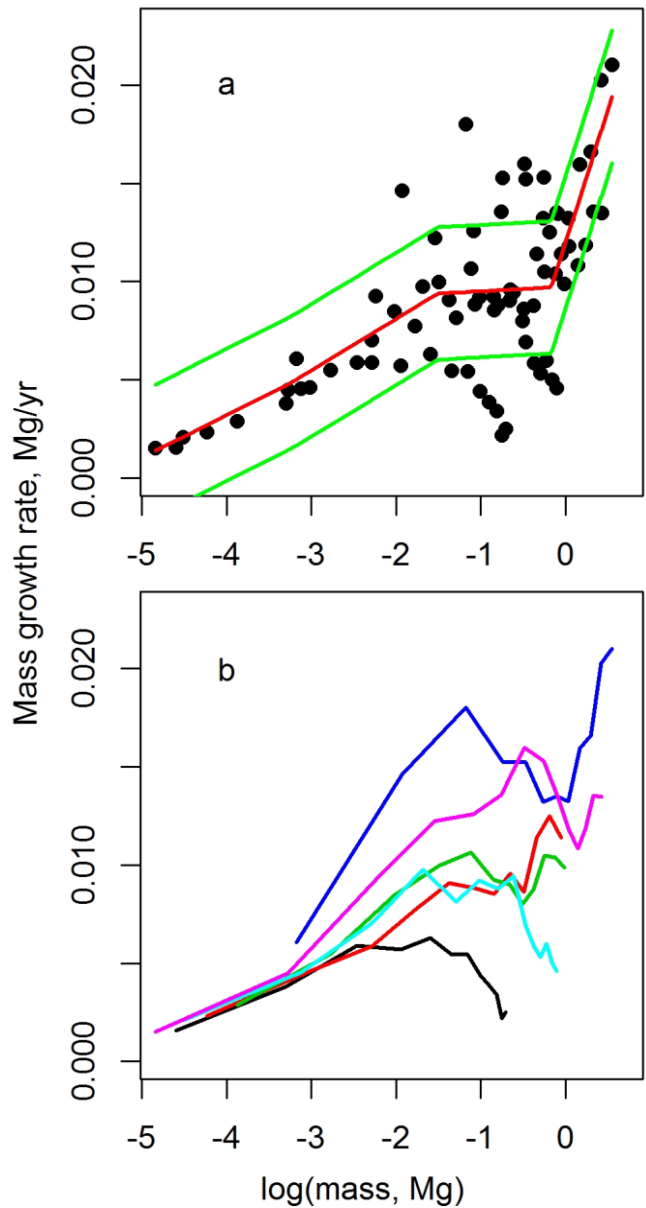

Figure 1 Mass growth data from von Guttenberg's Hinterberg Site Class I trees[5]. Panel 'a' shows a 4-bin piecewise linear regression through the data cloud, while panel 'b' shows the actual trajectories of growth of the individual trees

Stephenson et al.'s data are not timeseries of individual trees as we show here, but are isolated observations from many different trees. In effect, this is as if they had a random sample from a point cloud such as that in figure 1a, with single observations from a wide range of possible timeseries. It is unsurprising that a regression through such a cloud has an increasing trend, but it is simply false to infer that this trend applies to the individual trees. Stephenson et al.'s conclusions that rates of tree carbon accumulation increase continuously with tree size are therefore invalid.

The authors declare no competing financial interests.